\documentclass[prb,twocolumn,showpacs,preprintnumbers,amsmath,amssymb,superscriptaddress,aps]{revtex4-1}
\usepackage{amsmath,amssymb,amsfonts} 	
\usepackage{graphicx}
\usepackage{hhline}
\usepackage[utf8]{inputenc}

\begin{document}

\title{Theory and simulations of critical temperatures in CrI3 and other 2D materials: Easy-axis magnetic order and easy-plane Kosterlitz-Thouless transitions.}

\author{Thomas Olsen}
\email{tolsen@fysik.dtu.dk}
\affiliation{Computational Atomic-scale Materials Design (CAMD), Department of Physics, Technical University of Denmark, 2800 Kgs. Lyngby Denmark}

\begin{abstract}
The recent observations of ferromagnetic order in several two-dimensional (2D) materials have generated an enormous interest in the physical mechanisms underlying 2D magnetism. In the present prospective article we show that Density Functional Theory (DFT) combined with either classical Monte Carlo simulations or renormalized spin-wave theory can predict Curie temperatures for ferromagnetic insulators that are in quantitative agreement with experiment. The case of materials with in-plane anisotropy is then discussed and it is argued that finite size effects may lead to observable magnetic order in macroscopic samples even if long range magnetic order is forbidden by the Mermin-Wagner theorem. 
\end{abstract}
\maketitle

\section{Introduction}
The Mermin-Wagner theorem implies that two-dimensional (2D) materials cannot exhibit a spontaneously broken symmetry at finite temperatures.\cite{Studio2005} This means that long range magnetic order in 2D can only exist by virtue of magnetic anisotropy. More precisely, the continuous symmetry must be broken by the presence of an easy axis (typically out-of-plane in 2D materials), which breaks the spin rotational symmetry whereas an easy plane (typically the atomic plane of a 2D material) will yield a residual $O(2)$ symmetry that still prohibits (long range) magnetic order at finite temperatures.

In 2017, a monolayer of CrI$_3$ was shown to exhibit ferromagnetic order below 45 K\cite{Huang2017a} and CrI$_3$ thus comprises the first realization of magnetic order in 2D. The search for magnetism in this particular material was inspired by the presence of a strong out-of-plane easy axis (as required by the Mermin-Wagner theorem) in bulk CrI$_3$, which is a van der Waals bonded layered material. Since then, a few other materials have joined the family of 2D magnets. In particular, ferromagnetic order has been observed in Fe$_3$GeTe$_2$ below 130 K,\cite{Fei2018} evidence for room temperature ferromagnetism has been reported in MnSe$_2$\cite{doi:10.1021/acs.nanolett.8b00683} and VSe$_2$,\cite{Bonilla2018, Liu2018d} and FePS$_3$ has been shown to exhibit antiferromagnetic order below 118 K.\cite{Lee2016} In all cases, the magnetic order is driven by magnetic anisotropy that originates from spin-orbit interactions and is expected to be large for materials containing heavy elements. For example, CrI$_3$ and Fe$_3$GeTe$_2$ exhibit strong magnetic anisotropy due to the presence of I and Te atoms respectively. On the other hand,  MnSe$_2$ and VSe$_2$ are the only 2D materials to date that exhibit magnetic order at room temperature despite the lack of elements heavier than Se. This highlights that magnetic order in 2D materials is determined by a subtle interplay between magnetic anisotropy and exchange interactions and both effects may contribute to sizable critical temperatures.  Moreover, Fe$_3$GeTe$_2$, MnSe$_2$ and VSe$_2$ are itinerant ferromagnets and the microsocopic mechanism leading to magnetic order may be somewhat different than the cases of CrI$_3$ and NiPS$_3$, which are insulators. Providing estimates of critical temperatures in 2D materials from a given material composition thus remains a highly challenging problem.

It is not possible to calculate thermodynamic properties (such as critical temperatures) directly from first principles methods using present day techniques and one has to rely on certain approximate schemes to obtain quantitative predictions from theory. For insulators, the Heisenberg model provides an accurate account of the interactions between the localized spins that govern the properties of magnetic materials. The parameters entering the Heisenberg model can be calculated from Density Functional Theory (DFT), which thus constitutes an efficient framework for obtaining “first principles Heisenberg models”. Solving the Heisenberg model is, however, a non-trivial task. In particular, the importance of correlations in 2D implies that standard Weiss mean field theory\cite{Yosida1996} cannot yield faithful predictions and any approximation has to explicitly incorporate correlation. One approach is to calculate the non-interacting spin-wave spectrum (magnons), and then apply a temperature dependent renormalization scheme that take magnon interactions into account.\cite{Yosida1996, Gong2017b, Lado2017, Torelli2019} The magnetization can then be calculated from the (renormalized) spin-wave spectrum as a function of temperature and the critical temperature is obtained as the point of vanishing magnetization. A somewhat orthogonal approach is to neglect quantum effects completely and perform Monte Carlo (MC) simulations of the classical Heisenberg model.\cite{Torelli2019, Akturk2017, Yasuda2005, Kim2019} In contrast to renormalized spin-wave theory, the MC approach includes all correlations in the model, but does not take into account the quantum nature of spins. It has recently been shown that the renormalized spin-wave theory yields qualitatively wrong results in systems with large magnetic anisotropy and MC simulations appear to be the most appropriate choice for extracting critical temperatures from Heisenberg models in 2D. In particular, MC simulations based on a first principles Heisenberg model recently predicted Curie temperatures of 50 K and 24 K for CrI$_3$ and CrBr$_3$ respectively, which is in good agreement with the experimental values of 45 K and 27 K.\cite{Torelli2019c}

CrI$_3$, CrBr$_3$, FePS$_3$, and Fe$_3$GeTe$_2$ are reported to exhibit magnetic order of the Ising type meaning that the monolayers have strong out-of-plane easy-axes and first principles calculations indicate that the same is true for MnSe$_2$.\cite{Torelli2019c} In contrast, anomalous Hall conductivities and hysteresis loops have clearly shown that VSe$_2$ exhibits an easy-plane, which appears to be at odds with the Mermin-Wagner theorem and the magnetic properties of VSe$_2$ is currently debated. In particular, angle-resolved photoemission experiments\cite{Duvjir2018} as well as first principles calculations\cite{Wong2019} have shown that VSe$_2$ exhibits a paramagnetic charge density wave (CDW) at elevated temperatures,\cite{Duvjir2018, Wong2019} and the long range magnetic order reported in Ref. \onlinecite{Bonilla2018} has not yet been reproduced experimentally. In addition, Cr$_2$Ge$_2$Te$_6$20 and NiPS$_3$\cite{Kim2019} have been shown to exhibit easy-plane ferromagnetic and antiferromagnetic order respectively in bilayer structures, but both materials lack long range magnetic order in the monolayer limit as expected from the Mermin-Wagner theorem. The magnetic properties of VSe$_2$ are thus strongly complicated by the proximity of the ferromagnetic ground state with a paramagnetic charge density wave and the current understanding of the magnetic properties must be regarded as unresolved. However, from a fundamental point of view it seems hard to reconcile 2D magnetic order with clear experimental signatures of easy plane magnetization.

There is, however, a loophole for the Mermin-Wagner theorem that allow some amount of magnetization in easy-plane 2D magnets. If one considers the limit of infinite easy-plane anisotropy the spins effectively become confined to the atomic plane and the physics is captured by the so-called $XY$ model. The thermodynamic properties are then governed by Kosterlitz-Thouless (KT) physics and the spins exhibit algebraic correlations at all temperature below the KT phase transition. If that is the case there will be no characteristic length scale for correlations and any macroscopic sample will exhibit a finite magnetization due to finite size effects.\cite{Bramwell1994}

In this prospective, we will show how to calculate critical temperatures for 2D magnetic insulators using first principles Heisenberg models and classical Monte Carlo simulations. We then discuss the case of easy-plane magnetism and review the basic results from Kosterlitz-Thouless theory, which allow finite magnetization in macroscopic samples.

\section{The anisotropic Heisenberg model}
The Heisenberg model can be derived as an approximation to the many-body electronic Hamiltonian in a localized basis where only the spin degrees of are retained. In that framework the interactions between localized spins are referred to as direct exchange and arise as a consequence of Coulomb interactions and Pauli exclusion.\cite{Yosida1996} A rather different type of magnetic interactions may be derived by assuming strongly localized spins and include hybridization by second order perturbation theory. This will typically be mediated by non-magnetic ligand atoms and is referred to as superexchange.\cite{Anderson1950, Anderson1959} Remarkably, the superexchange interaction between localized spins has precisely the same form as the direct exchange interaction and can be included in the Heisenberg model  by a simple redefinition of the interactions parameters.
A rather general form of the Heisenberg model, which includes magnetic anisotropy can be written as
\begin{align}\label{eq:H}
H=-\frac{1}{2}\sum_{ij\alpha\beta}S_i^\alpha \mathcal{J}_{ij}^{\alpha\beta}S_i^\beta-\sum_{i\alpha} A_i^\alpha (S_i^\alpha)^2,
\end{align}
where $S_i^\alpha$ is the $\alpha$ component of the spin operator for the magnetic atom at site i and $\mathcal{J}_{ij}^{\alpha\beta}$ is a 3x3 interaction matrix that couple spin states at sites $i$ and $j$. The second sum contains single-ion anisotropy constants that determine the energy cost of a global spin rotation. In general the off-diagonal elements of $J_{ij}^{\alpha\beta}$ lead to Kitaev interactions\cite{Xu2018, Banerjee2016} and Dzyaloshinskii-Moriya interactions\cite{Heide2009, Koretsune2018, Liu2018} that may give rise to highly intriguing physical properties. However, the off-diagonal components are typically an order of magnitude smaller than the isotropic part\cite{Xu2018} and have marginal influence on the ordering temperatures. In the following we will thus stick to a simplified version where the off-diagonal components of $\mathcal{J}_{ij}^{\alpha\beta}$ is neglected and we assume in-plane isotropy such that  $J_{ij}^{xx}=J_{ij}^{yy}$ and $A_i^x=A_i^y$. The Heisenberg model can then be written as 
\begin{align}\label{eq:H_iso}
H=&-\frac{\widetilde J_{ij}}{2}\sum_{ij}\mathbf{S}_i\cdot\mathbf{S}_j-\frac{\widetilde J_{ij}^z}{2}\sum_{ij}\Big(2S_i^z S_j^z-S_i^x S_j^x-S_i^y S_j^y\Big)\notag\\
&-\sum_{i\alpha} A_i^\alpha (S_i^\alpha)^2,
\end{align}
where $A_i=A_i^z-A_i^x$, $\widetilde J_ij^z=(\mathcal{J}_{ij}^{zz}-\mathcal{J}_{ij}^{xx})/3$, and $\widetilde{J}_{ij}=Tr[\mathcal{J}_{ij}^{\alpha\beta}]/3$ is the isotropic part of the exchange tensor. Eq. \eqref{eq:H_iso} is then equivalent to Eq. \eqref{eq:H} except for a constant term given by $\sum_iA_i^xS_i(S_i+1)$, where $S_i$ is the maximum eigenvalue of $S_i^z$. In Eq. \eqref{eq:H_iso} we have explicitly split the exchange tensor into an isotropic and a traceless part. However, it is often more convenient to rewrite it in the form 
\begin{align}\label{eq:H_convenient}
H=-\frac{1}{2}\sum_{ij}J_ij\mathbf{S}_i\cdot\mathbf{S}_j-\frac{1}{2}\sum_{ij}\lambda_{ij}S_i^zS_j^z--\sum_iA_i(S_i)^2,
\end{align}
with $J_{ij}=\widetilde J_{ij}-\widetilde J_{ij}^z$ and $\lambda=3\widetilde J_{ij}^z$ This is the version of the Heisenberg Hamiltonian that will be used in the following. We note that different conventions for the parameters are used in the literature and it is always crucial to specify a convention when referring to a set of parameters.

\section{Heisenberg parameters from first principles}
Before delving into an analysis of Eq. \eqref{eq:H_convenient}, we will briefly show how to obtain the Heisenberg parameters  $J_{ij}$, $\lambda_{ij}$, and $A_i$ from first principles simulations.  It is in principle, possible to calculate the exchange coupling constants from microscopic expressions involving either an exchange integral in the case of direct exchange or from a combination of hopping matrix elements and on-site Coulomb interactions in the case of superexchange.\cite{Besbes2019, Wang2019} This approach is, however, problematic for several reasons: 1) A quantitative calculation has to be carried out in a localized basis set (for example Wannier functions\cite{Marzari2012}) and the results will inevitably depend on a specific choice of basis. 2) The exchange mechanism is often a combination of exchange and superexchange and there is no way of a priori determining how to balance the contributions in a quantitative treatment of the parameters. 3) Methods based on density functional theory (DFT) can only yield accurate total energies (if a good xc-functional is used), but “Kohn-Sham parameters” such as band energies, hopping parameters or exchange integrals do not comprise genuine physical properties.

A much better approach is based on an energy mapping analysis, where DFT total energies corresponding to different spin symmetries\cite{Grling1993, Grling2000} are mapped to the model (3) and used to extract the parameters.\cite{Xiang2013, Jacobsson2017, Kdderitzsch2002, Pajda2001, Olsen2017, Bose2010} This requires no knowledge of the underlying exchange mechanism and the accuracy is only limited by quality of the applied exchange-correlation functional. In particular, if we restrict ourselves to a single type of magnetic atom and nearest neighbor interactions only, a classical treatment of the Heisenberg model leads to\cite{Torelli2019}
\begin{align}
J=&\frac{E_{AFM}^\parallel-E_{FM}^\parallel}{N_{AFM}S^2},\label{eq:J}\\
\lambda=&\frac{E_{FM}^\parallel-E_{FM}^\perp-E_{AFM}^\parallel+E_{AFM}^\perp}{N_{AFM}S^2},\label{eq:lambda}\\
A=&\frac{(E_{FM}^\parallel-E_{FM}^\perp)\delta_--(E_{AFM}^\parallel-E_{AFM}^\perp)\delta_+}{2S^2},\label{eq:A}
\end{align}
where $E_{FM}$ and $E_{AFM}$ are the energies of the ferromagnetic and an antiferromagnetic configuration respectively. The superscripts denotes whether the magnetization is in-plane ($\parallel$) or out-of-plane ($\perp$) and $\delta_\pm=1\pm N_{FM}/N_{AFM}$ where $N_{FM}$ and $N_{AFM}$ is the number of nearest neighbors with aligned and anti-aligned spins respectively in the anti-ferromagnetic configuration. In the case of a bipartite lattice of magnetic atoms there is a fully anti-ferromagnetic configuration with with $N_{FM}=0$ and $\delta_\pm=1$. However, for-non-bipartite lattices (a triangular lattice for example) there is no natural choice of the anti-ferromagnetic state and one has to choose a frustrated configuration with $N_{FM}\neq0$. Typically, the Heisenberg parameters are on the order of meV or less and it is crucial to use the same basis sets for all configurations in order to obtain well converged results.  This approach is easily generalized to yield additional Heisenberg parameters such as next-nearest and third-nearest neighbor interactions, but for insulators these are typically much smaller than the nearest neighbor interactions, which usually govern the critical temperatures. In general, however, one has to check whether it is sufficient to restrict the analysis to nearest neighbor interactions.

If the ground state is antiferromagnetic it may not be obvious, which value to use for $S$ in the energy mapping analysis. However, a ferromagnetic configuration without spin-orbit coupling is bound to yield a half integer for any insulator and that defines the value of S to be used in the Heisenberg model. We note that DFT codes often provide an estimate of local magnetic moments based on an integrated spin-density in some vicinity of the atomic nuclei, but this comprises a rather arbitrary measure and should not be used for quantitative analysis. Moreover, spin-orbit effects may yield a non-integer value of the magnetic moments (in unit of Bohr magnetons per cell) in a ferromagnetic configuration, which is at odds with the Heisenberg model that assumes a half-integer spin in the ferromagnetic ground state. This is due to inter-band spin mixing, which is not captured by the Heisenberg model and such effects cannot be captured by a model like Eq. \eqref{eq:H_convenient}. While it would be highly interesting to extend the Heisenberg model to include multiple band effects this is a rather small effect that is unlikely to affect the prediction of critical temperatures. For example, in CrI$_3$ the inclusion of spin-orbit coupling leads to an LDA magnetic moment per Cr ion of 3.01 Bohr magnetons, which is very close to the nominal value of 3.0. Finally, the inclusion of spin-orbit coupling that is needed to obtain the four DFT energies $E_{FM/AFM}^{\perp/\parallel}$  introduces certain subtleties in the calculations. In the case of CrI$_3$ the ground state is ferromagnetic with a strong out-of-plane easy axis and from this state $E_{FM}^\perp$ is obtained. However, the ferromagnetic state with in-plane magnetization comprises a saddle point in spin-configuration space and any unconstrained calculation will converge towards the state with out-of-plane magnetization. To resolve this one can either perform a constrained DFT calculation or include spin-orbit corrections non-selfconsistently. 

It should be emphasized that the energy mapping analysis leading to Eqs. \eqref{eq:J}-\eqref{eq:A} was based on the classical Heisenberg model. For a ferromagnetic system with easy-axis anisotropy the ground state energy coincides with the classical energy. But for an anti-ferromagnetic system, the classical ground state energy only provides an upper estimate of the true ground state energy. For example, for a square lattice with nearest neighbor interactions and no anisotropy the ground state energy per site is approximately $2JS^2(1+0.158/S)$\cite{Yosida1996} whereas the classical energy is simply $2JS^2$ ($J$ is negative). If one assumes that DFT provides the correct ground state energy, the classical energy mapping analysis will lead to an overestimation of the exchange coupling constant. This problem also pertains to ferromagnetic systems since the antiferromagnetic state that enters the analysis is presumably correctly described in DFT and therefore does not correspond to the classical energy assumed in Eqs. \eqref{eq:J}-\eqref{eq:A}. In principle it is straightforward to include quantum corrections in the analysis for any isotropic Heisenberg model, but the situation becomes somewhat more complicated when anisotropy is introduced. For example, a ferromagnetic Heisenberg model with out of plane easy axis has a lowest and a highest eigenstate respectively that must correspond to $E_{FM}^\perp$ and $E_{AFM}^\parallel$ obtained from DFT. But the energies $E_{FM}^\parallel$ and $E_{AFM}^\perp$ cannot be mapped onto any eigenstate of the Heisenberg model and it is not obvious, how to proceed with the quantum mechanical energy mapping in this case.

\begin{table}[t!]
\centering
\begin{tabular}{|c|c|c|c|c|c|} 
 \hline
  & LDA & LDA+U & PBE & PBE+U & PBEsol \\ 
 \hline
 $\quad J\quad$ & 1.28 (1.28) & 2.97 & 2.09 & 3.81 & 2.14 \\ 
 \hline
 $A$ & 0.22 (0.19) & 0.024 & 0.16 & -0.009 & 0.17 \\
 \hline
 $\lambda$ & 0.16 (0.17) & 0.25 & 0.13 & 0.27 & 0.15 \\
 \hline
 $\Delta$ & 1.18 (1.16) & 1.17 & 0.92 & 1.18 & 1.00\\
 \hline
\end{tabular}
\caption{Exchange and anisotropy constants of a monolayer CrI$_3$ calculated with a few different functionals. All values are in meV. Spin-orbit coupling was included non-selfconsistently except for the values in brackets, which were obtained with selfconsistent spin-orbit coupling. LDA+U and PBE+U was performed with U=3.5 eV}
\label{tab}
\end{table}
The accuracy of commonly applied exchange-correlation functionals for exchange and anisotropy constant has not yet been completely clarified, but there has been a few studies on the performance of hybrids\cite{Kdderitzsch2002} and LDA+U/PBE+U for three-dimensional bulk systems.\cite{Olsen2017} It was found that for the anti-ferromagnets NiO and MnO, LDA+U and PBE+U were able to provide good agreement with experimental values for nearest and next nearest neighbor interactions if the “right” value of U was chosen. In Tab. 1, we compare the Heisenberg parameters for CrI$_3$ calculated with LDA, LDA+U, PBE, PBE+U and PBEsol, we have used a value of U=3.5 eV for LDA+U and PBE+U. All calculations were performed with the electronic structure code GPAW, which is based on plane waves and the projector-augmented wave method.\cite{Enkovaara2010a, Olsen2016a, Larsen2017} All structures were relaxed with the given functional. It is observed that the predicted exchange constants may differ by more than a factor of three depending on the applied functional. In addition the anisotropy constant is strongly dependent on the whether or not an on-site Coulomb repulsion U is introduced. However, the spin-wave gap $\Delta=A(2S-1)+\lambda N_{nn}S$ (to be introduced below), which largely determines the spin-orbit mediated magnetic properties is not very sensitive to the functional. This signifies that the spin-orbit effects are rather insensitive to the xc-functional. Including an onsite U term in the calculation merely transfers the single-ion anisotropy energy into the anisotropic exchange energy. However, the large span in exchange constants indicates that first principles predictions of magnetic properties such as critical temperatures may be off by a factor of three and there is a strong need for a systematic study of the accuracy of exchange-correlation functionals for exchange constants.

\section{Easy-axis order}
Once a first principles Heisenberg Hamiltonian has been obtained through the energy mapping analysis we are left with the pertinent problem of solving it. In three-dimensional magnets a rough estimate of the critical temperature can be obtained from Weiss mean field theory. In 2D, however, this approach is bound to fail since there is no reference to the dimensionality of the problem and finite critical temperatures will be predicted even in the absence of anisotropy. In 2D the effect of anisotropy must be included on equal footing with the exchange interactions. In the presence of an easy-axis (here chosen as the z-direction) a rigorous quantum mechanical analysis can be carried out by introducing the Holstein-Primakoff transformation, which replaces the spin operators in Eq. \eqref{eq:H_convenient} by bosonic operators that create or annihilate spin wave excitations (magnons). For a single type of magnetic atom the transformation is 
\begin{align}
S_i^-=&\sqrt{2S}a_i^\dag \sqrt{1-\frac{(a_i^\dag a_i)}{2S}},\\
S_i^+=&\sqrt{2S}\sqrt{1-\frac{(a_i^\dag a_i)}{2S}}a_i ,\\
S_i^z=&S-a_i^\dag a_i,
\end{align}
with $S_i^\pm=S_i^x\pm iS_i^y$. The bosonic commutation relations for $a_i$ and $a_i^\dag$ imply that the commutation relations for the spin operators are fulfilled. In order to proceed, the square roots must be Taylor expanded and one obtains
\begin{align}\label{eq:H_exp}
H=E_0+H_2+H_4+H_6+\ldots,
\end{align}
where $E_0$ is zeroth order in raising and annihilation operators, $H_2$ is second order, $H_4$ is fourth order and so forth. In Eq. \eqref{eq:H_exp} we assume each term to be normal ordered such that all annihilation operators are to the right. This implies that $E_0$ is simply the ground state energy, $H_2$ determines the spectrum of single magnon excitations, $H_4$ gives rise to two-magnon interactions, and $H_6$ gives rise to three-magnon interactions. It should be noted that the anisotropy parameters only enter $E_0$, $H_2$, and $H_4$ and a truncation at fourth order thus includes anisotropy exactly. 

If one is interested in the properties at low temperatures it may be assumed that two-magnon excitations are rare and the spectrum can be obtained from $H_2$, which can be solved directly by a Fourier transform. The spectrum then becomes\cite{Torelli2019}
\begin{align}\label{eq:eps}
\varepsilon_n(\mathbf{q})=\varepsilon_n^0(\mathbf{q})+A(2S-1)+\lambda SN_nn,
\end{align}
where $n$ denotes a band index (the range is equal to the number of magnetic atoms in the unit cell) and $\varepsilon_n^0(\mathbf{q})$ is the spectrum without anisotropy, which satisfies $\varepsilon_{n_0}^0(\mathbf{0})=0$ for the lowest band $n_0$. The term $\Delta_0\equiv A(2S-1)+\lambda SN_{nn}$ is therefore the spin-wave gap and magnetic order in 2D relies on $\Delta_0\neq0$. It should be noted that for $S=1/2$ the single-ion anisotropy term cannot open a gap and anisotropic exchange is required for magnetic order.
\begin{figure}
  \includegraphics[width=\linewidth]{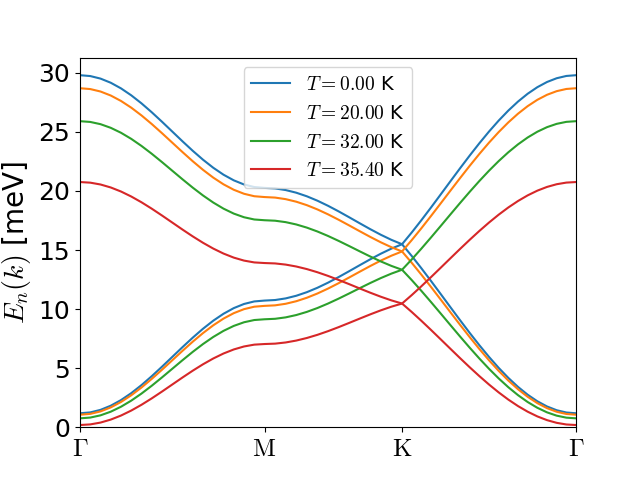}
  \caption{Renormalized spin-wave spectrum of CrI$_3$ calculated at different temperatures.}
  \label{fig:sw}
\end{figure}

\begin{figure}
  \includegraphics[width=\linewidth]{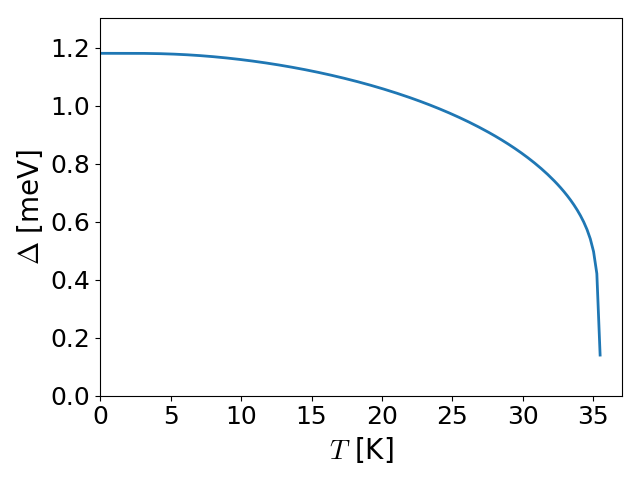}
  \caption{Temperature dependent gap of CrI$_3$ calculated from renormalized spin-wave theory Eq. \eqref{eq:delta}. The gap closes at $T=35$ K,
which signals a phase transition with loss of magnetic order.}
  \label{fig:delta}
\end{figure}

At finite temperatures the spin-wave approximation breaks down due to the presence of thermally excited magnons. The magnon interactions may be included through a mean field approximation of the fourth order term, which leads to a temperature dependent (renormalized) spin-wave spectrum.\cite{Yosida1996, Tyablikov2013, Gong2017b} In particular, the gap is effectively decreased by temperature effects and for a single site in the unit cell it may be written as\cite{Torelli2019}
\begin{align}\label{eq:delta}
\Delta=\Delta_0-4A\langle n\rangle-\lambda N_{nn}\langle n\rangle, 
\end{align}
where $\langle n\rangle$ is the Bose distribution of magnons averaged over the Brillouin zone. Since the Bose distribution depends on temperature as well as the magnon spectrum one has to calculate the gap and spectrum self-consistently. In Fig. \ref{fig:sw}, we show the renormalized spectrum of CrI$_3$ calculated at different temperature using the parameters of PBE+U from Tab. \ref{tab} It is clear that the gap decreases as the temperature is increased. The Bose distribution of magnons will diverge at the point where the gap vanishes and this signals a phase transition in the present approximations. In Fig. \ref{fig:delta}, we show the spin-wave gap as a function of temperature and it is seen to vanish at $T_c= 35$ K. This appears to be in good agreement with the experimental value of 45 K. However, the agreement might be fortuitous and certainly depends on our choice of PBE+U values for the Heisenberg parameters.

\begin{figure}
  \includegraphics[width=\linewidth]{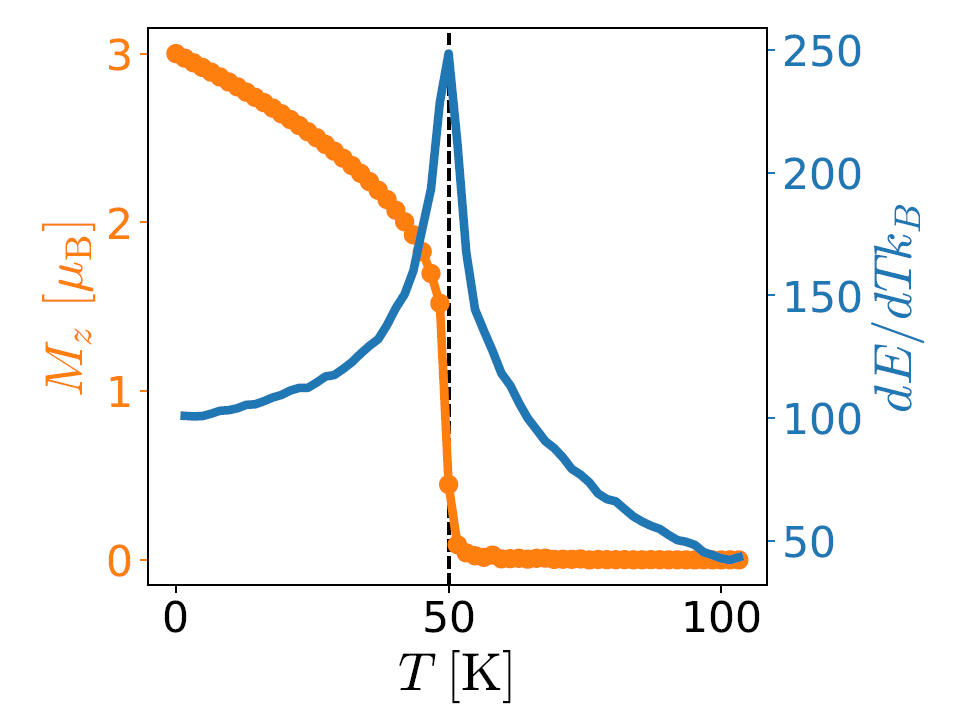}
  \caption{Magnetization per Cr atom ($M_z$ in unit of Bohr magnetons) and heat capacity ($dE/dT$) of a monolayer CrI$_3$ obtained from
Monte Carlo simulations. Both quantities show a clear phase transition at a $T=50$ K where the magnetic order is lost.}
  \label{fig:mc}
\end{figure}

\begin{figure}
  \includegraphics[width=\linewidth]{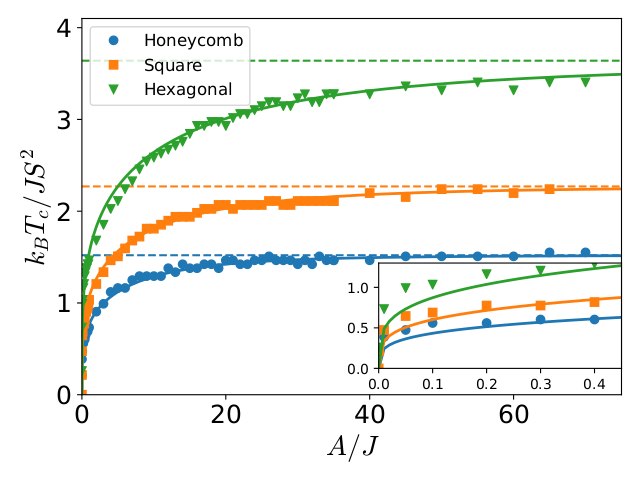}
  \caption{Critical temperature as a function of single-ion anisotropy calculated from Monte Carlo simulations of Honeycomb, square, and hexagonal (triangular) lattices. The solid lines are fitted functions given by Eq \eqref{eq:T_c}. Figure from Ref. \onlinecite{Torelli2019} ©IOP Publishing. Reproduced with permission. All rights reserved.}
  \label{fig:T}
\end{figure}

It is far from clear that the mean field approach is a good approximation close to the critical temperature where a high density of magnons is expected. In fact, it is not even obvious that multiple-magnon interactions (terms beyond fourth order in the Hamiltonian) can safely be neglected. On the other hand, quantum effects tend to be quenched by thermal fluctuations at elevated temperatures and for the sole purpose of evaluating critical temperatures a purely classical approach might be expected to work well. The average energy and magnetization of the classical Hamiltonian \eqref{eq:H_convenient} can be obtained straightforwardly from Monte Carlo simulations and in Fig. \ref{fig:mc} we show an example of such calculations; again using the Heisenberg parameters obtained with PBE+U. The magnetization decreases monotonously with increasing temperature and drops abruptly to zero at $T = 50$ K. Similarly, the heat capacity has a sharp peak at $T = 50$ K, which thus signals a phase transition where magnetic order is lost. The critical temperatures resulting from such calculations can be tabulated for different lattices and values of anisotropy parameters. In Fig. \ref{fig:T} we display such a compilation for honeycomb, square and triangular lattices in case of $\lambda=0$. In the limit of $A\rightarrow\infty$ the in-plane spin components are quenched and the critical temperature approaches that of the corresponding Ising model with coupling parameter $J$. This is an exact condition, which is naturally obtained in a classical treatment, but breaks down in renormalized spin-wave theory.\cite{Torelli2019}

As it turn out all the MC calculations can be fitted to the function\cite{Torelli2019}
\begin{align}\label{eq:T_c}
T_c = S^2T_c^{Ising}\tanh^{1/4}\bigg[\frac{6}{N_{nn}}\log\Big(1+\gamma\frac{\Delta_0}{J(2S-1)}\Big)\bigg],
\end{align}
where $\gamma=0.033$, $N_{nn}$ is the number of nearest neighbors, and  $T_c^{Ising}$ is the critical temperature of the corresponding Ising model, which is given by 1.52, 2.27 and 3.64 in units of $J/k_B$ for the honeycomb, square and triangular lattices respectively. This expression is useful for high throughput computational screening of ferromagnetic compounds, since it only requires three Heisenberg parameters that are easily obtained from DFT as explained above. Such an approach has recently been applied to the Computational 2D Materials Database (C2DB)\cite{Pandey2018} where 3712 2D materials was screened for magnetic properties and yielded a prediction of 17 novel stable 2D ferromagnetic materials.\cite{Torelli2019c} It should also be straightforward to conduct a screening study on experimental databases like the Inorganic Crystal Structure Database (ICSD)\cite{Allmann2007} and the Crystallography Open Database (COD)\cite{Graulis2011} by applying a measure that identifies exfoliable materials from three-dimensional parent compounds.\cite{Mounet2018, Larsen2019} 

\section{Easy-plane order}
The above analysis assumes an easy-axis that coincides with the out-of-plane direction. If the ground state has a component of magnetization in the plane of the material and in-plane isotropy is assumed, the rotational freedom prohibits long range magnetic order due to the Mermin-Wagner theorem. Mathematically, the above analysis would then result in a negative spin-wave gap that implies the instability of a state with out-of-plane magnetization. For high throughput calculations the sign of the spin-wave gap can thus be used as a descriptor that determines if long-range order is possible at finite temperatures.

\begin{figure}
  \includegraphics[width=\linewidth]{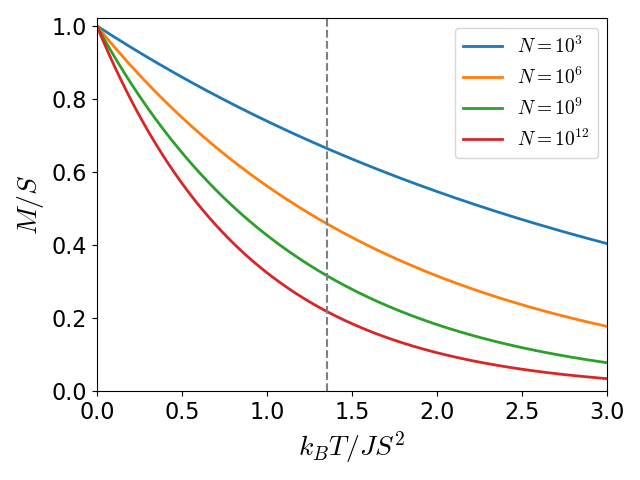}
  \caption{The magnetization in units of $S$ as a function of temperature in the $HXY$ model. The dashed vertical line indicates the KT transition above which the spins become disordered due to unbound vortex/anti-vortex pairs.}
  \label{fig:M}
\end{figure}

Although long range order is forbidden for easy-plane magnets in 2D, finite size effects may give rise to a macroscopic magnetization if the anisotropy is large. For example, in the limit of $A\rightarrow-\infty$ the $S^z$ components of the spins become quenched and the model \eqref{eq:H_convenient} effectively reduces to the $XY$ model given by 
\begin{align}\label{eq:XY}
H_{XY}=-\frac{1}{2}\sum_{ij}J_{ij} (S_i^x S_j^x+S_i^y S_j^y).
\end{align}
Kosterlitz and Thouless have shown that this model exhibits critical behavior for all temperatures below a certain temperature $T_{KT}$, meaning that spin correlations decay with a power law dependence on distance. This implies that long range order may be observed in any macroscopic sample although the order strictly speaking vanishes for an infinite system. Assuming small spin deflections such that $\mathbf{S}_i\cdot\mathbf{S}_j=S^2\cos(\theta)\approx S^2 (1-\theta ^2/2)$ in Eq. \eqref{eq:XY}, yields the harmonic $XY$ model ($HXY$) and a classical analysis then shows that the magnetization of a finite sample is given by\cite{Jos1977}
\begin{align}\label{eq:M}
M=\bigg(\frac{1}{2N}\bigg)^{k_B T/8\pi J},
\end{align}
where $N$ is the number of sites in the system. However, the analysis neglects the possibility of spin vortices that effectively renormalize the magnetization. The energy of a single vortex diverges, but bound vortex/anti-vortex pairs have a finite energy and constitute fundamental excitations of the system at low energies. At the Kosterlitz-Thouless transition temperature, the vortex/anti-vortex pairs unbind and the system becomes disordered. In Fig. \ref{fig:M}, we show the magnetization \eqref{eq:M} as a function of temperature for different values of $N$. We also indicate $T_{KT}^{HXY}=1.351 J/k_B$ above which the magnetic order is destroyed by unbound vortex/anti-vortex pairs. As shown by Bramwell and Holdsworth, the KT temperature is largely unaltered by finite size effects although strictly speaking it only marks a true phase transition in the infinite system.\cite{Bramwell1994} This implies that the magnetization is lost above the KT temperature and we may take the critical temperature for ordering as $T_c\approx T_{KT}^{HXY}$. Only the magnitude of the magnetization below $T_{KT}^{HXY}$ will depend on the size of a sample and will vanish in the limit of an infinite system. The decay of magnetization with system size is, however, very slow. For example, a sample with an area of 1 $\mu$m$^2$ and a distance between magnetic atoms of 3 Å, the magnetization is reduced by 60 {\%} at the KT transition compare to the value at $T=0$. For a 1 mm$^2$ sample the reduction will be 80 {\%} and a macroscopic magnetization will thus be prominent at any temperature below 
$T_{KT}^{HXY}$. The phase transition in the $XY$ model (without the harmonic approximation) is a bit lower ($T_{KT}=0.898J/k_B$), but the conclusion remains the same.

In real materials the condition of $A\rightarrow-\infty$ will of course not be satisfied, but part of the Kosterlitz-Thouless behavior may be expected in easy-plane 2D magnets with finite anisotropy. Macroscopic samples of 2D materials with an easy-plane may thus show magnetic order due to finite size effects even if the Mermin-Wagner theorem prohibits long-range magnetic order. In particular the easy-plane 2D magnets, Cr$_2$Ge$_22$Te$_6$\cite{Gong2017b}, NiPS$_3$\cite{Kim2019} and VSe$_2$\cite{Bonilla2018} could exhibit magnetism of the Kosterlitz-Thouless type described above. Only VSe$_2$ have been reported to show macroscopic magnetization at finite temperatures though and those measurements have subsequently been questioned.\cite{Wong2019} Nevertheless, the subtle correlations and intricate magnetic frustration reported for VSe$_2$ could indicate that this material exhibits rich magnetic phenomena and the presence of spin vortices could hinder the direct measurements of magnetization and perhaps make it strongly dependent on the precise conditions under which the measurements are carried out. In the case of Cr$_2$Ge$_2$Te$_6$ the KT transitions is expected at $\sim J/k_B$ which is roughly 75 K,\cite{Gong2017b, Xu2018} but the persistence of in-plane magnetic order is hindered by the exceedingly small anisotropy in this material.

\section{Outlook}
In this prospective article we have shown how to obtain accurate predictions for critical temperatures of 2D ferromagnetic insulators using DFT combined with either renormalized spin-wave theory or classical Monte Carlo simulations. Both methods agree well with experiments for the cases of and CrI$_3$ and CrBr$_3$, but the renormalized spin-wave method is expected to break down in highly anisotropic systems. We also presented a universal, but simple analytical expression for the critical temperature that were fitted to MC simulations and only depends on the Heisenberg parameters as well the number of nearest neighbors in a given material. Thus, the critical temperature of any 2D ferromagnetic insulator can be estimated from just four DFT calculations. 

A major caveat of this method is the fact that insulating magnetic materials often exhibit strongly correlated physics and the calculated Heisenberg parameters can be rather sensitive to the chosen DFT functional. As such DFT must be regarded as inaccurate for the prediction of Heisenberg parameters in these systems. At least there is a need for more systematic assessment of the performance of various functionals for the prediction of magnetic properties. Another – perhaps more fundamental – deficiency of the method is the energy mapping analysis, which is based on the classical Heisenberg model. The classical Monte Carlo simulations of Curie temperatures are justified because the thermal fluctuations quench the quantum fluctuations in the vicinity of the critical temperature, but the parameters are calculated from the ground state at zero K, and should be mapped to the quantum mechanical Heisenberg model. Without anisotropy it is straightforward to obtain the quantum corrected value of the nearest neighbor exchange coupling constants. It is however challenging to generalize that analysis to the case where anisotropy is included, but we hope that the challenge will be taken up in near future such that a universal and rigorous energy mapping scheme can be defined.

In the present prospective, we have focused on the calculation of transition temperatures for ferromagnetic order in 2D materials. A natural next step is then to extend the analysis to anti-ferromagnetic order in 2D, which has been observed in FePS$_3$\cite{Lee2016, Wang2016} and NiPS$_3$.\cite{Kim2019, Wildes2015} The calculation of Heisenberg parameters and Monte Carlo simulations would  proceed exactly as in the ferromagnetic case and for any bipartite lattice (for example square or honeycomb) the spin-wave analysis can be carried out by means of a Bogoliubov transformation.\cite{Yosida1996} However, for non-bipartite lattices (for example triangular) anti-ferromagnetic exchange coupling constants will give rise to geometric frustration and the magnetic phase diagram may become extremely rich with several competing phases for a given set of Heisenberg parameters.\cite{Chernyshev2009, Maksimov2019} Even in the absence of anisotropy the classical ground state is non-collinear and the mechanism by which anisotropy opens a gap is much more complicated than in the ferromagnetic case. A thorough analysis of the effect of anisotropic exchange on the magnetic properties of triangular lattices has recently been carried out,\cite{Maksimov2019} but an analysis of the effect of single-ion anisotropy in non-bipartite lattices with anti-ferromagnetic exchange coupling still seems to be lacking.

The most important remaining question is how to deal with itinerant magnets in 2D. Most of the reported high-temperature 2D magnets like FePS$_3$,\cite{Lee2016, Wang2016} VSe$_2$\cite{Bonilla2018} and Fe$_3$GeTe$_2$\cite{Fei2018} are indeed metallic, but it is not at all clear if the Heisenberg model \eqref{eq:H_convenient} can provide even a qualitatively correct description in those cases. Naively, one may expect that the Heisenberg model could provide a decent approach if the exchange coupling constants can be converged with respect to distance. However, there are several issues with such an approach; it is – for example – not clear what value to use for the spin $S$. Alternatively, one may try to apply Stoner theory of magnetism,\cite{Yosida1996} which comprises a more natural starting point for itinerant magnets, but there is no simple way to obtain critical temperatures in that case and the theory requires inclusion of a parametric Hubbard U, that measures on-site Coulomb repulsion between electrons. The prediction of thermodynamic properties for itinerant magnets thus poses the biggest and most important theoretical challenge in the field at the moment.

2D materials comprise an extremely versatile class of compounds. There are endless possibilities of optimizing particular properties by bottom up design of van der Waals heterostructures composed of 2D materials. In the case of magnetic 2D materials, van der Waals heterostructures have already been shown to yield new magnetic properties. For example, multilayers of CrI$_3$ exhibit anti-ferromagnetic interlayer coupling and the coercive field for interlayer alignment range from 0.65 T to 1.84 T depending on the number of layers.\cite{Huang2017a, Sivadas2018, Jiang2019} This implies that the magnetic properties of multilayer structures of CrI$_3$ can be tuned simply by varying the number of layers. The anti-ferromagnetic structure of multilayers also implies that multilayers can act as spin valves with an interlayer conductance that may change by several orders of magnitude depending on the inter-layer magnetic state.\cite{Cardoso2018, Song2018, Wang2018, Song2019} An additional convenient feature of 2D materials is the ease at which they can be electrostatically gated and for 2D magnets the magnetic properties may change dramatically under the influence of a gate voltage.\cite{Huang2018, Jiang2018, Morell2018} Finally, bilayers of CrI$_3$ have symmetry-allowed nearest neighbor Dzyaloshinskii-Moriya interactions that could give rise to Skyrmions\cite{Liu2018, Liu2018e} or perhaps topological magnons.\cite{Mook2014, Pershoguba2018} For other materials, such as Cr$_2$Ge$_2$Te$_6$ and NiPS$_3$, the bulk intra-layer magnetic order (ferromagnetic and anti-ferromagnetic respectively) is preserved down to a bilayer, but vanishes in the monolayer limit. This is due to the easy-plane nature of the anisotropy, by which the long-range order is lacking for a monolayer, whereas even the bilayer structure bypasses the Mermin-Wagner theorem by weak interlayer magnetic interactions. It remains to be seen, however, if a remnant of Kosterlitz-Thouless physics can be observed in monolayers of such materials.


%

\end{document}